\documentstyle[prd,aps,amsfonts]{revtex}
\begin{document}
\def\bh{${\Bbb R}^2\times {\Bbb S}^2\>$}
\draft
\title{ Twisted spinors on black holes}
\author{Yu. P. Goncharov}
\address{Theoretical Group, Experimental Physics Department, State Technical
         University, Sankt-Petersburg 195251, Russia}
\date{Submitted 1 April 1999}
\maketitle
\begin{abstract}
  We remark that the standard black hole topology admits twisted
configurations of spinor field due to existence of the twisted spinor bundles
and analyse them using the Schwarzschild black hole as an example.
This is physically linked with the natural
presence of Dirac monopoles on black holes and entails
marked modification of the Hawking radiation for spinor particles.
\end{abstract}

1.  A few years ago there appeared an interest in studying
topologically inequivalent configurations (TICs) of
various fields on the 4D black holes
\cite{{Gon9456},{GF967},{Gon978},{Gon979}}
since TICs might give marked additional
contributions to the quantum effects in the 4D black hole physics, for
instance, such as the Hawking radiation \cite{GF967} and also might help
to solve the problem of statistical substantiation of the black hole entropy
\cite{Gon978}. So far, however, only TICs of complex scalar field
have been studied more or less on the
Schwarzschild (SW), Reissner-Nordstr\"om (RN) and Kerr black holes. The next
physically important case is the one of spinor fields. In the present paper
we start studying twisted TICs of spinor field in the form convenient to
physical applications, restricting here ourselves to the framework of the SW
black hole geometry for the sake of simplicity.

We write down the black hole
metric under discussion (using the ordinary set of local coordinates
$t,r,\vartheta,\varphi$) in the form
$$ds^2=g_{\mu\nu}dx^\mu\otimes dx^\nu\equiv
adt^2-a^{-1}dr^2-r^2(d\vartheta^2+\sin^2\vartheta d\varphi^2) \eqno(1)$$
with $a=1-r_g/r$, $r_g=2M$, where $M$
is a black hole mass. Besides we have
$|g|=|\det(g_{\mu\nu})|=(r^2\sin\vartheta)^2$
and $r_g\leq r<\infty$, $0\leq\vartheta<\pi$,
$0\leq\varphi<2\pi$.

  Throughout the paper we employ the system of units with $\hbar=c=G=1$,
unless explicitly stated otherwise. Finally, we shall denote $L_2(F)$ the set
of the modulo square integrable complex functions on any manifold $F$
furnished with an integration measure.

2.  The existence of spinor field TICs on black holes
follows from the fact that over the standard black hole topology \bh\,
there exists
only one Spin-structure [conforming to the group
Spin(1,3)= SL$(2,{\Bbb C})$].
Referring for the exact definition of Spin-structure to
Refs.\cite{{81},{89}}, we here only
note that the number of inequivalent Spin-structures for manifold $M$ is equal
to the one of elements in $H^1(M,{\Bbb Z}_2)$, the first cohomology group
of $M$ with coefficients in ${\Bbb Z}_2$. In our case
$H^1($\bh$,{\Bbb Z}_2)=
H^1({\Bbb S}^2,{\Bbb Z}_2)$ which is
equal to 0 and thus there exists the only (trivial) Spin-structure.

On the other hand, the nonisomorphic
complex line bundles over $M$ are classified by
the elements in $H^2(M,{\Bbb Z})$,
the second cohomology group of $M$ with coefficients in {\Bbb{Z}}
\cite{Gon9456}, and in our case this group is equal to
$H^2({\Bbb S}^2,{\Bbb Z})={\Bbb Z}$ and, consequently, the number of
complex line bundles is countable. As a result,
each complex line bundle can be characterized
by an integer $n\in{\Bbb Z}$ which in what follows will be called its
Chern number.

Under this situation, if denoting
$S(M)$ the only standard spinor bundle over $M$ = \bh and $\xi$ the complex
line bundle with Chern number $n$, we can construct tensorial product
$S(M)\otimes\xi$.
As is known \cite{GI}, over any noncompact spacetime the bundle
$S(M)$ is trivial and, accordingly, the Chern number of 4-dimensional vector
bundle $S(M)\otimes\xi$ is equal to $n$ as well. Under the circumstances we
obtain the {\it twisted Dirac operator}
${\cal D}: S(M)\otimes\xi\to S(M)\otimes\xi$, so the wave equation for
corresponding spinors $\psi$ (with a mass $\mu_0$) as sections of the bundle
$S(M)\otimes\xi$ may look as follows
$${\cal D}\psi=\mu_0\psi,\>\eqno(2)$$
and we can call (standard) spinors corresponding to $n=0$ (trivial complex
line bundle $\xi$) {\it untwisted} while the rest of the spinors with $n\ne0$
should be referred to as {\it twisted}.

 From general considerations \cite{{81},{89},{Bes87}} the explicit form of
the operator ${\cal D}$ in local coordinates $x^\mu$ on a $2k$-dimensional
(pseudo)riemannian manifold can be written as follows
$${\cal D}=i\nabla_\mu\equiv i\gamma^aE_a^\mu(\partial_\mu-
\frac{1}{2}\omega_{\mu ab}\gamma^a\gamma^b-ieA_\mu),\>a < b ,\>\eqno(3)$$
where $A=A_\mu dx^\mu$ is a connection in the bundle $\xi$ and the forms
$\omega_{ab}=\omega_{\mu ab}dx^\mu$ obey the Cartan structure equations
$de^a=\omega^a_{\ b}\wedge e^b$ with exterior derivative $d$, while the
orthonormal basis $e^a=e^a_\mu dx^\mu$ in cotangent bundle and
dual basis $E_a=E^\mu_a\partial_\mu$ in tangent bundle are connected by the
relations $e^a(E_b)=\delta^a_b$. At last, matrices $\gamma^a$ represent
the Clifford algebra of
the corresponding quadratic form in ${\Bbb C}^{2^k}$. Below we shall deal only
with 2D euclidean case (quadratic form $Q_2=x_0^2+x_1^2$) or
with 4D lorentzian case (quadratic form $Q_{1,3}=x_0^2-x_1^2-x_2^2-x_3^2$).
For the latter case we take the following choice for $\gamma^a$
$$\gamma^0=\pmatrix{1&0\cr 0&-1\cr}\,,
\gamma^b=\pmatrix{0&\sigma_b\cr-\sigma_b&0\cr}\,,
b= 1,2,3\>, \eqno(4)$$
where $\sigma_b$ denote the ordinary Pauli matrices.
It should be noted that, in lorentzian case, Greek indices $\mu,\nu,...$
are raised and lowered with $g_{\mu\nu}$ of (1) or its inverse $g^{\mu\nu}$
and Latin indices $a,b,...$ are raised and lowered by
$\eta_{ab}=\eta^{ab}$= diag(1,-1,-1,-1),
so that $e^a_\mu e^b_\nu g^{\mu\nu}=\eta^{ab}$,
$E^\mu_aE^\nu_bg_{\mu\nu}=\eta_{ab}$ and so on.

Using the fact that all the mentioned bundles $S(M)\otimes\xi$ can be
trivialized over the chart of local coordinates
$(t,r,\vartheta,\varphi)$ covering almost the whole manifold
\bh , we can concretize the wave equation (2) on the given chart for
TIC $\psi$ with the Chern number $n\in{\Bbb Z}$ in the case of metric (1).
Namely, we can put $e^0=\sqrt{a}dt$, $e^1=dr/\sqrt{a}$,
$e^2=rd\vartheta$, $e^3=r\sin{\vartheta}d\varphi$ and, accordingly,
$E_0=\partial_t/\sqrt{a}$, $E_1=\sqrt{a}\partial_r$,
$E_2=\partial_\vartheta/r$, $E_3=\partial_\varphi/(r\sin{\vartheta})$.
This entails
$$\omega_{01}=-\frac{1}{2}\frac{da}{dr}dt,
\omega_{12}=-\sqrt{a}d\vartheta,
\omega_{13}=-\sqrt{a}\sin{\vartheta}d\varphi,
\omega_{23}=-\cos{\vartheta}d\varphi.\>\eqno(5)$$
As for the connection $A_\mu$ in bundle $\xi$ then the suitable one was found
in Refs.\cite{Gon9456} and is
$A= A_\mu dx^\mu=-\frac{n}{e}\cos\vartheta d\varphi\>.$
Then the curvature of the bundle $\xi$ is $F=dA=
\frac{n}{e}\sin\vartheta d\vartheta\wedge d\varphi$. We can further introduce
the Hodge star operator on 2-forms $F$ of any $k$-dimensional
(pseudo)riemannian manifold
$B$ provided with a (pseudo)riemannian metric $g_{\mu\nu}$ by the relation
(see, e. g., Ref.\cite{Bes87})
$$F\wedge\ast F=(g^{\mu\alpha}g^{\nu\beta}-g^{\mu\beta}g^{\nu\alpha})
F_{\mu\nu}F_{\alpha\beta}
\sqrt{|g|}\,dx^1\wedge dx^2\cdots\wedge dx^k \eqno(6)$$
in local coordinates $x^\mu$. In the case of the metric (1) this yields
$\ast F= \frac{n}{er^2}dt\wedge dr\>,$
and integrating over the surface $t= const$, $r=const$ with topology
${\Bbb S}^2$ gives rise to
the Dirac charge quantization condition
$$\int_{S^2} F=4\pi\frac{n}{e}=4\pi q \eqno(7)$$
with magnetic charge $q$,
so we can identify the coupling constant $e$ with electric charge.
Besides, the Maxwell equations $dF=0$, $d\ast F =0$ are clearly fulfilled
with the exterior differential $d=\partial_t dt+\partial_r dr+
\partial_\vartheta d\vartheta+\partial_\varphi d\varphi$ in coordinates
$t,r,\vartheta,\varphi$.
We come to the same conlusion that in the case
of TICs of complex scalar field \cite{{Gon9456},{GF967},{Gon978},{Gon979}}:
the Dirac magnetic $U(1)$-monopoles naturally live on the black holes as
connections in complex line bundles  and hence physically the appearance of
TICs for spinor field should be
obliged to the natural presence of Dirac monopoles on black hole and due to
the interaction with them the spinor field splits into TICs.
Also it should be emphasized that the total
(internal) magnetic charge $Q_m$ of black hole which
should be considered as the one summed up over all the monopoles remains
equal to zero because
$$Q_m=\frac{1}{e}\sum\limits_{n\in{\Bbb{Z}}}\,n=0\>.\eqno(8)$$

Returning to the Eq. (2), we can see that with taking into account
all the above it has the form
$$[i\gamma^0\frac{1}{\sqrt{a}}(\partial_t-
\frac{1}{2}\omega_{t01}\gamma^0\gamma^1)+
i\gamma^1\sqrt{a}\partial_r+
i\gamma^2\frac{1}{r}(\partial_\vartheta-
\frac{1}{2}\omega_{\vartheta12}\gamma^1\gamma^2)+$$
$$i\gamma^3\frac{1}{r\sin{\vartheta}}(\partial_\varphi-
\frac{1}{2}\omega_{\varphi13}\gamma^1\gamma^3-
\frac{1}{2}\omega_{\varphi23}\gamma^2\gamma^3+
in\cos{\vartheta})]\psi = \mu_0\psi  \>.     \eqno(9)$$
After a simple matrix algebra computation with using
(4) and (5) we find that Eq.(9) can be rewritten as
$$\pmatrix{A&B\cr -B&-A\cr}\pmatrix{\psi_1\cr\psi_2\cr}=
\mu_0\pmatrix{\psi_1\cr\psi_2\cr}\>, \eqno(10)$$
with the operators
$$A=\frac{i}{\sqrt{a}}\partial_t,
B=i\sigma_1B_1+\frac{1}{r}B_2 \>, \eqno(11)$$
where, in turn,
$$B_1=\frac{1}{2}\frac{d\sqrt{a}}{dr}+\sqrt{a}\partial_r+\frac{\sqrt{a}}{r},
B_2=i\sigma_2\partial_\vartheta+i\sigma_3\frac{1}{\sin{\vartheta}}
(\partial_\varphi-\frac{1}{2}\sigma_2\sigma_3\cos{\vartheta}+
in\cos{\vartheta})\>.\eqno(12) $$
Now we can use the ansatz
$\psi_1=e^{i\omega t}r^{-1}F_1(r)\Phi(\vartheta,\varphi)$,
$\psi_2=e^{i\omega t}r^{-1}F_2(r)\sigma_1\Phi(\vartheta,\varphi)$ with a
2D spinor $\Phi=\pmatrix{\Phi_1\cr\Phi_2}$ in order with the help (10)--(12)
to get
$$  (B_1+\frac{1}{r}D_n)\psi_1=i(\mu_0-c)\sigma_1\psi_2,$$
$$  (B_1+\frac{1}{r}D_n)\psi_2=-i(\mu_0+c)\sigma_1\psi_1\>\eqno(13)  $$
with $c=\frac{1}{\sqrt{a}}\omega$ and $D_n=-i\sigma_1B_2$.
It is natural to take $\Phi$ as an eigenspinor of the operator $D_n$ and
noting that $\sigma_1D_n=-D_n\sigma_1$ we can from (13) obtain the system
$$\sqrt{a}\partial_rF_1+
\left(\frac{1}{2}\frac{d\sqrt{a}}{dr}+\frac{\lambda}{r}\right)F_1=
i(\mu_0-c)F_2,$$
$$\sqrt{a}\partial_rF_2+
\left(\frac{1}{2}\frac{d\sqrt{a}}{dr}-\frac{\lambda}{r}\right)F_2=
-i(\mu_0+c)F_1 \>\eqno(14)$$
with an eigenvalue $\lambda$ of the operator $D_n$. We should, therefore,
explore the equation $D_n\Phi=\lambda\Phi$.

3. As is not complicated to see, the operator $D_n$ has the form (3) with
$\gamma^0=-i\sigma_1\sigma_2$, $\gamma^1=-i\sigma_1\sigma_3$,
$e^0=d\vartheta$, $e^1=\sin{\vartheta}d\varphi$,
$\omega_{01}=\cos{\vartheta}d\varphi$,
$A_\mu dx^\mu= -\frac{n}{e}\cos{\vartheta}d\varphi$,
i. e., it corresponds to the abovementioned quadratic form $Q_2$ and this is
just twisted (euclidean) Dirac operator on the unit sphere and the conforming
complex line bundle is the restriction of bundle $\xi$ on the unit sphere.
Again simple matrix algebra computation results in
$D_n=\pmatrix{D_{1n}&D_{2n}\cr-D_{2n}&-D_{1n}\cr}$ with
$D_{1n}=i(\partial_\vartheta+\frac{1}{2}\cot{\vartheta})$,
$D_{2n}=-\frac{1}{\sin{\vartheta}}(\partial_\varphi+in\cot{\vartheta})$.
Then it is easy to see that the equation $D_n\Phi=\lambda\Phi$ can be
transformed into the one
$$\pmatrix{0&D^-_n\cr D^+_n&0\cr}\Phi_0=\lambda\Phi_0,
\Phi_0=\pmatrix{\Phi_+\cr\Phi_-\cr}\>, \eqno(15)$$
where $D_n^\pm=D_{1n}\pm D_{2n}=
i[\partial_\vartheta+(\frac{1}{2}\mp n)\cot{\vartheta}]
\mp\frac{1}{\sin{\vartheta}}\partial_\varphi$, $\Phi_{\pm}=\Phi_1\pm\Phi_2$.
From here it follows that $D_n^-D_n^+\Phi_+=\lambda^2\Phi_+$,
$D_n^+D_n^-\Phi_-=\lambda^2\Phi_-$ or, with employing the ansatz
$ \Phi_\pm=P_\pm(\vartheta)e^{-im'\varphi}$, in explicit form
$$\left[\partial^2_\vartheta+\cot{\vartheta}\partial_\vartheta-
\frac{m'^2+(n\mp1/2)^2-2m'(n\mp1/2)\cos{\vartheta}}{\sin^2{\vartheta}}
\right]P_\pm(\vartheta)=$$
$$\left(\frac{1}{4}-n^2-\lambda^2\right)P_\pm(\vartheta)
\>. \eqno(16) $$
It is known \cite{Vil91} that differential operators of the left-hand side
in (16) have eigenfunctions in the interval $0\le\vartheta\le\pi$, which are
finite at $\vartheta=0,\pi$, only for eigenvalues $-k(k+1)$, where $k$ is
positive integer or half-integer simultaneously with $m',n'=n\pm1/2$ while the
multiplicity of such an eigenvalue is equal to $2k+1$. In our
case we have that $n'=n\pm1/2$ is half-integer because the Chern
number $n\in{\Bbb Z}$. As a result, we should put $m'=m-1/2$ with
an integer $m$ and then $|m'|\le k=l+1/2$ with a positive integer $l$ and,
accordingly, $1/4-n^2-\lambda^2=-k(k+1)$ which entails (denoting $\lambda=
\sqrt{(l+1)^2-n^2})$ that spectrum of $D_n$ consists of the numbers
$\pm\lambda$
with multiplicity $2k+1=2(l+1)$ of each one. Besides, it is clear that
under the circumstances $-l\le m\le l+1, l\ge|n|$. This just reflects 
the fact that
from general considerations \cite{{81},{89},{Bes87}} the spectrum of twisted
euclidean Dirac operator on even-diemensional manifold is symmetric with
respect the origin.
The corresponding eigenfunctions
$P_\pm(\vartheta)=P^k_{m'n'}(\cos\vartheta)$ of the above operators
can be chosen in miscellaneous forms (see, e. g.,
Ref. \cite{Vil91})
with the orthogonality relation at $n'$ fixed
$$\int\limits_0^\pi\,{P^{*k}_{m'n'}}(\cos\vartheta)
P^{k'}_{m'' n'}(\cos\vartheta)
\sin\vartheta d\vartheta={2\over2k+1}\delta_{kk'}
\delta_{m'm''}\>,\eqno(17)$$
where (*) signifies complex conjugation. As a consequence, we come to the
conclusion that spinor $\Phi_0$ of (15) can be chosen in the form
$\Phi_0=\pmatrix{C_1P^k_{m'n-1/2}\cr C_2P^k_{m'n+1/2}\cr}e^{-im'\varphi}$
with some constants $C_{1,2}$. Now we can employ the relations \cite{Vil91}
$$-\partial_\vartheta P^k_{m'n'}\pm\left(n'\cot{\vartheta}-
\frac{m'}{\sin{\vartheta}}\right)P^k_{m'n'}=
-i\sqrt{k(k+1)-n'(n'\pm1)}P^k_{m'n'\pm1} \eqno(18) $$
holding true for functions $P^k_{m'n'}$ to establish that $C_1=C_2=C$
corresponds to eigenvalue $\lambda$ while $C_1=-C_2=C$ conforms to
$-\lambda$. Thus, the eigenspinors $\Phi=\pmatrix{\Phi_1\cr\Phi_2\cr}$ of
the operator $D_n$ can be written as follows
$$\Phi_{\pm\lambda}=\frac{C}{2}\pmatrix{P^k_{m'n-1/2}\pm P^k_{m'n+1/2}\cr
P^k_{m'n-1/2}\mp P^k_{m'n+1/2}\cr}e^{-im'\varphi}\>, \eqno(19) $$
where the coefficient $C$ may be defined from the normalization condition
$$\int\limits_0^\pi\,\int\limits_0^{2\pi}(|\Phi_1|^2+|\Phi_2|^2)
\sin\vartheta d\vartheta d\varphi=1\>    \eqno(20)$$
with using the relation (17) that yields $C=\sqrt{\frac{l+1}{\pi}}$. These
spinors form an orthonormal basis in $L_2^2({\Bbb S}^2)$. Finally, it should
be noted that the given spinors can be expressed through the {\it monopole
spherical harmonics} $Y^l_{mn}(\vartheta,\varphi)=
P^l_{mn}(\cos{\vartheta})e^{-im\varphi}$
which naturally arise when considering twisted TICs of complex scalar field
\cite{{Gon9456},{GF967},{Gon978},{Gon979}} but we shall not need it here.

4. As follows from the above, when quantizing twisted TICs of spinors 
we can take the set of spinors
$$\psi_{\pm\lambda}=\frac{1}{\sqrt{2\pi\omega}}
e^{i\omega t}r^{-1}\pmatrix{F_1(r,\pm\lambda)
\Phi_{\pm\lambda}\cr
F_2(r,\pm\lambda)\sigma_1\Phi_{\pm\lambda}\cr}\> \eqno(21)$$
as a basis in $L_2^4$(\bh) and realize
the procedure of quantizing, as usual, by expanding in the modes (21)
$$\psi=\sum\limits_{\pm\lambda}\sum\limits_{l=|n|}^\infty
\sum\limits_{m=-l}^{l+1}
\int\limits_{\mu_0}^\infty\,d\omega
(a^-_{\omega nlm}\psi_{\lambda}+
b^+_{\omega nlm}{\psi_{-\lambda}})\,,$$
$$\overline{\psi}=
\sum\limits_{\pm\lambda}\sum\limits_{l=|n|}^\infty\sum\limits_{m=-l}^{l+1}
\int\limits_{\mu_0}^\infty\,d\omega
(a^+_{\omega nlm}\overline{\psi}_{\lambda}+
b^-_{\omega nlm}\overline{\psi}_{-\lambda})\,,\eqno(22)$$
where $\overline{\psi}=\gamma^0\psi^{\dag}$ is the adjont spinor and ($\dag$)
stands for hermitian conjugation.
As a result, the operators
$a^{\pm}_{\omega nlm}$, $b^{\pm}_{\omega nlm}$ of (22) should be
interpreted as the
creation and annihilation ones for spinor particle in the
gravitational field of the black hole and in the field of monopole with
Chern number $n$. As to the
functions $F_{1,2}(r,\pm\lambda)$ of (21) then in accordance with Eqs. (14) 
we can get
the second order equations for them in the form
$$a\partial_ra\partial_rF_{1,2}+a\left[\sqrt{a}\partial_r\left(\frac{1}{2}
\frac{d\sqrt{a}}{dr}\pm\frac{\lambda}{r}\right)+
\frac{1}{4}\left(\frac{d\sqrt{a}}{dr}\right)^2-\frac{\lambda^2}{r^2}\right]
F_{1,2}=\left(a\mu_0^2-\omega^2\right)F_{1,2} \>. \eqno(23)$$
By replacing
$r^*=r+r_g\ln(r/r_g-1)$
and by going to the dimensionless quantities $x=r^*/M,y=r/M, k= \omega M$
the equations (23) can be rewritten in the Schr\"odinger-like equation form
$$\frac{d^2}{dx^2}E_{1,2}+[k^2-(\mu_0M)^2]E_{1,2}=
V_{1,2}(x,\lambda)E_{1,2}    \eqno(24) $$
with $E_{1,2}=E_{1,2}(x,k,\lambda)=F_\pm(Mx),
F_\pm(r^*)=F_{1,2}[r(r^*)]$ while the potentials $V_{1,2}$ are given by
$$V_{1,2}(x,\lambda)=\frac{1}{4y^4(x)}+
\left[\frac{1}{y^4(x)}\mp\frac{\lambda}{y^2(x)}
\sqrt{1-\frac{2}{y(x)}}+
\frac{\lambda^2}{y^2(x)}\right]
\left[1-\frac{2}{y(x)}\right]
-\frac{2}{y(x)}(\mu_0M)^2
\>,    \eqno(25)$$
where $y(x)$ is a function reverse to $x(y)=y+2\ln(0.5y-1)$,
so $y(x)$ is the one-to-one correspondence between ($-\infty,\infty$) and
$(2,\infty)$.

Let us for the sake of simplicity restrict ourselves
to the massless spinors ($\mu_0=0$). Then,
as can be seen, when $x\to+\infty$, $V_{1,2}\to0$ and at $x\to-\infty$,
$V_{1,2}\to1/64$. This allows us to pose the scattering problem on the
whole $x$-axis for Eq. (24) at $k>0$
$$E_{1,2}^+\sim\cases{e^{ikx}+
\ s_{12}^{(1,2)}e^{-ikx}+\frac{1}{64k^2},
&$x\rightarrow-\infty$,\cr
 s_{11}^{(1,2)}e^{ikx},&$x\rightarrow+\infty$,\cr} $$
$$E_{1,2}^-\sim\cases{s_{22}^{(1,2)}
e^{-ikx}+\frac{1}{64k^2},
&$x\rightarrow-\infty$,\cr
e^{-ikx}+s_{21}^{(1,2)}e^{ikx},&$x
\rightarrow+\infty$\cr}\eqno(26)$$
with $S$-matrices $\{s_{ij}^{(1,2)}=s_{ij}^{(1,2)}(k,\lambda)\}$. Then
by virtue of (14) one can obtain the equality
$$s_{11}^{(1)}(k,\lambda)=-s_{11}^{(2)}(k,\lambda)
                                   \>.\eqno(27)  $$
  Having obtained these relations, one
can speak about the Hawking radiation process for any TIC of spinor
field on black holes. Actually, one can notice that Eq. (2) corresponds
to the lagrangian
$${\cal L}=\frac{i}{2}|g|^{1/2}\left[\overline{\psi}\gamma^\mu\nabla_\mu\psi-
(\nabla_\mu\overline{\psi})\gamma^\mu\psi-\mu_0\overline{\psi}\psi\right]
\>, \eqno(28)$$
and one can use the energy-momentum tensor
for TIC with the Chern number $n$ conforming to the lagrangian (28)
$$T_{\mu\nu}=\frac{i}{4}\left[\overline{\psi}\gamma_\mu\nabla_\nu\psi+
\overline{\psi}\gamma_\nu\nabla_\mu\psi-
(\nabla_\mu\overline{\psi})\gamma_\nu\psi-
(\nabla_\nu\overline{\psi})\gamma_\mu\psi\right] \>,\eqno(29) $$
to get, according to the standard
prescription (see, e. g., Ref. \cite{Gal86}) with
employing (20) and (27), the luminosity $L(n)$ with respect to the Hawking
radiation for TIC with the Chern number $n$ (in usual units)
$$L(n)=\lim_{r\to\infty}\,\int\limits_{S^2}\,
<0|T_{tr}|0>d\sigma=
A\sum\limits_{\pm\lambda}\sum\limits_{l=|n|}^\infty2(l+1)
\int\limits_0^\infty\,\frac{|s^{(1)}_{11}(k,\lambda)|^2}{e^{8\pi k}+1}dk
 \eqno(30)$$
with the vacuum expectation value $<0|T_{tr}|0>$ and the surface element
$d\sigma=r^2\sin\vartheta d\vartheta\wedge d\varphi$ while
$A=\frac{c^5}{GM}\left(\frac{c\hbar}{G}\right)^{1/2}
\approx0.125728\cdot10^{55}\,
{\rm{erg\cdot s^{-1}}}\cdot M^{-1}$ ($M$ in g).

We can interpret $L(n)$ as an additional contribution to the
Hawking radiation due to the additional spinor particles leaving black
hole because of the interaction with monopoles. Under this situation,
for the total luminosity $L$ of black hole with respect
to the Hawking radiation concerning the spinor field to be obtained,
one should sum up over all $n$, i. e.
$$L=\sum\limits_{n\in{\Bbb{Z}}}\,L(n)=L(0)+
2\sum\limits_{n=1}^\infty\,L(n)\,, \eqno(31)$$
since $L(-n)=L(n)$.

As a result, we can expect marked increase of Hawking radiation from
black holes for spinor particles. But for to get an exact value of this
increase one should apply numerical methods, so long as the scattering
problem for general equation (24) does not admit any exact solution and
is complicated enough for consideration --- the potentials
$V_{1,2}(x,\lambda)$ of (25) are
given in an implicit form.
One can remark that, for instance,
the similar increase for complex scalar field can amount to 17 \% of total
(summed up over all the TICs) luminosity \cite{GF967}. 

5. It is clear that the most general case is 
the Kerr-Newman black hole one but the equations here will
be more complicated so that we shall consider them elsewhere. 

    The work was supported in part by the Russian Foundation for
Basic Research (grant No. 98-02-18380-a) and by GRACENAS (grant No.
6-18-1997).

\end{document}